# Non Linear Susceptibility from High DC Field Torque Magnetometry

B.S. Shivaram

Department of Physics, University of Virginia, Charlottesville, VA. 22901

ABSTRACT

Torque magnetometry is a convenient technique to measure the magnetic properties of anisotropic materials. Advances in micromachining and the availability of robust materials with which such magnetometers can be fabricated has made them reliable even in adverse conditions such as very high magnetic fields and both high and very low temperatures. In most applications with such magnetometers the measured torque signals are used to arrive at the linear magnetic susceptibilities only. In this short note we extend torque magnetometry to measure nonlinear susceptibilities and illustrate our methods with representative data on the heavy fermion compound $UPt_3$.

The ready availability of stable and defect free materials such as single crystal silicon and quartz and advances in their micromachining have enabled the technique of torque magnetometry to be adopted on a wider scale[1]. The other competing technique used to obtain magnetization of samples, SQUID based magnetometry, also commonly employed and superior in certain ways, places several constraints on the experimental situation. SQUID magnetometry is quantitative due to the fundamental principle of flux quantization. It does not require a calibration sample if the area in which the flux is contained is precisely known. Torque magnetometry lagged in this regard for a while but has caught up with advances in microlithography. By placing a calibration loop on the magnetometer body encompassing the area where the sample is mounted - a known linearly dependent magnetic moment can be created by passing a measured amount of current through the loop to counterbalance the torque signal from the sample[2]. The sensitivity and stability of the signal is also enhanced when this feedback principle is employed making torque measurements superior to SQUID based techniques in high magnetic field measurements. Torque based methods also have the merits that liquid helium need not be employed whenever there is no need for measurements down to low temperatures. The deflection produced by the torque can be sensed in a variety of ways but most conveniently through capacitance detection when high magnetic fields are employed.

Given the aforementioned advantages of torque based methods to obtain magnetization in high dc magnetic fields it is worth investigating its use beyond the extraction of the linear susceptibility. In a recent work Kasahara et. al. have performed torque magnetization measurements with the field rotated in the basal plane of a tetragonal crystal $BaFe_2(As_{1-x}P_x)_2$ and find a two-fold symmetry instead of an in-plane isotropy expected for the basal plane[3]. Such a nematic state can arise in principle from anisotropies in higher order susceptibilities[4]. Rotational symmetry breaking and two-fold nematic order has also been detected in the heavy fermion compound $URu_2Si_2$ with torque[5]. In addition, the anisotropy of the higher order power dependences of the torque on the magnetic field ($H^4$ and $H^6$) in single crystal $TbAl_2$ have also been investigated[6]. While the relevance of the non-linear susceptibilities to the anisotropic dependence of the magnetization has been noted in the above three cited references neither of them provide a recipe for the extraction of the non-linear susceptibilities along specific crystal directions. Given the large current interest in understanding various types of exotic magnetic order in a variety of materials[7,8,9] the precise determination of the higher order magnetic susceptibilities is likely to grow as an experimental area. In fact in recent work we have shown that there are fundamental and universal relationships between linear and non-linear susceptibilities in a large variety of magnetic systems[10]. This adds to the need to develop methods to measure higher order components of the susceptibility tensor.

The leading order nonlinear dc susceptibility, $\chi_3$, has been studied in a number of magnetic systems[11,12,13,14,15,16,17,18,19,20] in recent years. In all of the reported work the methods employed are SQUID based with the exception of the measurements on $PrOs_4Sb_{12}$ where a Faraday balance was used[21,22]. In this paper we describe how torque magnetometry can also be used to extract similarly the nonlinear susceptibilities in a convenient manner.

Consider an anisotropic crystalline sample, with axes labeled 'c' and 'a' placed under an applied magnetic field H. The field H makes an angle θ with the c-axis as shown in fig. 1. The total torque τ on the sample will be

$$\vec{\tau} = \vec{m} \times \vec{H} = \mu_0 V(M_c H_x - M_a H_z) \qquad (1)$$

where $M_c$ and $M_a$ are projections of the magnetization M and $H_x$ and $H_z$ are the magnetic field components (z is taken along c-axis of the crystal). If measurements of the torque are performed for the magnetic field oriented close to the two orthogonal crystal axes then one can write:

$$\vec{\tau}_1 = (\vec{m} \times \vec{H})_{c-axis} = \mu_0 V(M_c H_x - M_b H_z) \qquad (2)$$

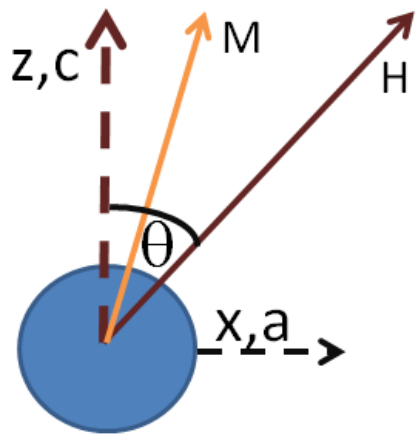

Figure 1: Schematic of an anisotropic crystalline sample placed in a magnetic field oriented at an angle θ with respect to the c-axis of the crystal. The coordinate system is fixed to the crystal.

and

$$\vec{\tau}_2 = (\vec{m} \times \vec{H})_{a-axis} = \mu_0 V(M_b H_x - M_c H_z) \qquad (3)$$

Taking the difference in the torque signals between the two orthogonal orientations one obtains:

$$\begin{aligned}\vec{\tau}_2 - \vec{\tau}_1 &= \mu_0 V[(M_b + M_c)H_x - (M_c + M_b)H_z] \\ &= \mu_0 V(M_b - M_c)H(Sin\theta + Cos\theta)\end{aligned} \qquad (4)$$

Adding the two torque signals yields:

$$\begin{aligned}\vec{\tau}_2 + \vec{\tau}_1 &= \mu_0 V[(M_b + M_c)H_x - (M_c + M_b)H_z] \\ &= \mu_0 V(M_b + M_c)H(Sin\theta - Cos\theta)\end{aligned} \qquad (5)$$

Rearranging terms the two components of the magnetization $M_a$ and $M_c$ can be written separately as:

$$M_a = \left[\frac{1}{2H}\right]\left[\frac{\vec{\tau}_2 + \vec{\tau}_1}{(Sin\theta - Cos\theta)} + \frac{\vec{\tau}_2 - \vec{\tau}_1}{(Sin\theta + Cos\theta)}\right] \qquad (6)$$

$$M_c = \left[\frac{1}{2H}\right]\left[\frac{\vec{\tau}_2 + \vec{\tau}_1}{(Sin\theta - Cos\theta)} - \frac{\vec{\tau}_2 - \vec{\tau}_1}{(Sin\theta + Cos\theta)}\right] \qquad (7)$$

For non-ferromagnetic metals we can expand these two magnetizations:

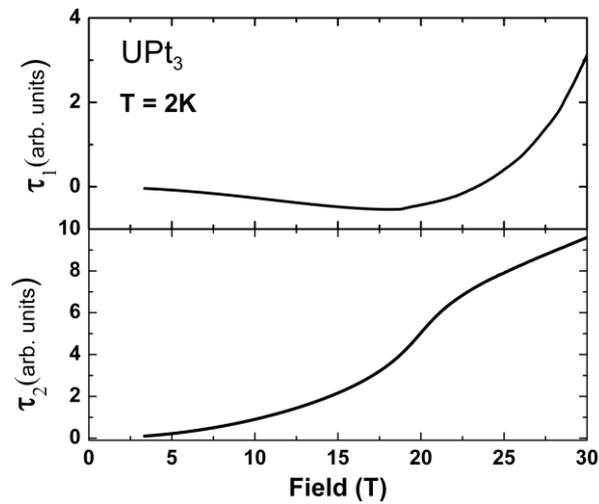

*Figure 2: Shows the torque signals, $\tau_1$ and $\tau_2$, obtained from a single crystal sample of UPt$_3$ when the magnetic field is applied about 1 degree away from the c-axis (top panel) and the a-axis (bottom panel) of the crystal.*

$$M_c = \chi_{1c}H_z + \chi_{3c}H^3_z + \chi_{5c}H^5_z \quad \text{with } H_z = HCos\theta \qquad (8)$$
$$M_a = \chi_{1a}H_x + \chi_{3a}H^3_x + \chi_{5a}H^5_x \quad \text{with } H_x = HSin\theta \qquad (9)$$

where $\chi_1$, $\chi_3$ and $\chi_5$ are the linear, third order and fifth order susceptibilities respectively. In writing eqns.(8) and (9) we have assumed that demagnetization effects are not important. We have also ignored the tensor nature of the susceptibilities $\chi_3$ and $\chi_5$ since the angle $\theta$ in our experiments is less than a degree and we are measuring only the diagonal components. If it turns out that there are off-diagonal components that are anomalously large compared to the diagonal parts then the above simple analysis can breakdown. The torque signals entering the above equations can be sensed in a variety of ways and most conveniently in high fields via capacitance measurements as stated earlier.

In figure 2 we show the raw data - labeled $\tau_1$ and $\tau_2$ - the capacitance torque signals for two orthogonal orientations of the magnetic field for a sample of $UPt_3$ a compound with hexagonal crystal symmetry. The magnetometer was constructed from single crystal silicon and a capacitance bridge was employed to measure the deflection due to the torque. An interesting aspect in $UPt_3$ is the occurrence of a metamagnetic transition characterized by a sudden increase in the magnetization at a critical field beyond which the growth in the magnetization is largely reduced or halted[23]. This transition is seen only when the field is applied parallel to the basal plane of the hexagonal crystal while the response in the orthogonal direction is similar to that of an ordinary paramagnet. To extract this well known metamagnetic behavior in $UPt_3$ from the two torque signals shown in fig. 1 we use the deconvolution procedure described by equations (6) and (7). The calculated values of the magnetization for the two directions are shown in fig.3 - the metamagnetic behavior is clearly seen in the main panel.

To proceed further and extract the nonlinear susceptibilities we can rewrite eqn.(9) as:

$$M_a / H = \chi_{1a} + \chi_{3a}H^2 + \chi_{5a}H^4 \qquad (10)$$

The measured values of M (arbitrary units) shown in fig. 2 have been plotted as per eqn.(10) in fig.3. The intercept on this plot is proportional to the linear susceptibility and the coefficients of the linear and quadratic terms in a least squares fit yields the third order and fifth order susceptibilities respectively. The results shown in fig.3 indicate that the fifth order susceptibility is significantly positive in UPt$_3$ for fields parallel to the basal plane. In this analysis we confine ourselves to a maximum field of 16 tesla. In terms of the square of the applied field (the horizontal axis in fig. 3) it corresponds to one order of magnitude higher than that used in our previous work where only the third order susceptibility $\chi_3$ was extracted[10].

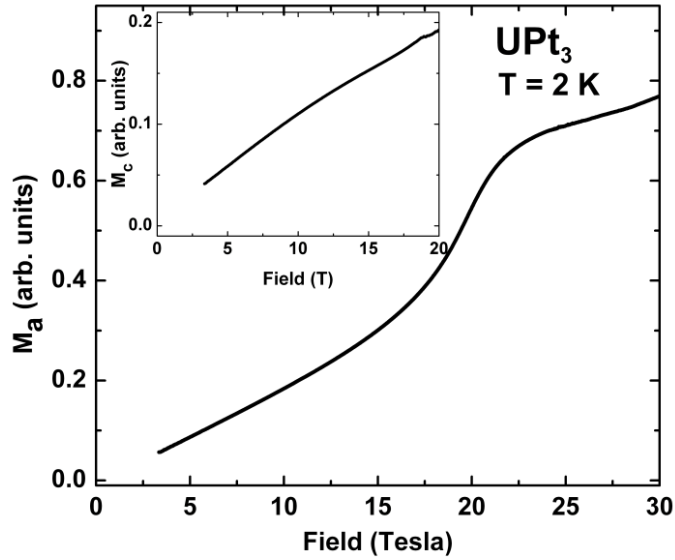

*Figure 3: Shows the magnetization along the a-axis obtained after deconvoluting the torque capacitance signals in the two orthogonal directions $\tau_1$ and $\tau_2$ shown in fig.2. The metamagnetic transition at 20 T is clearly seen. The inset shows the magnetization along the perpendicular direction (H||c-axis).*

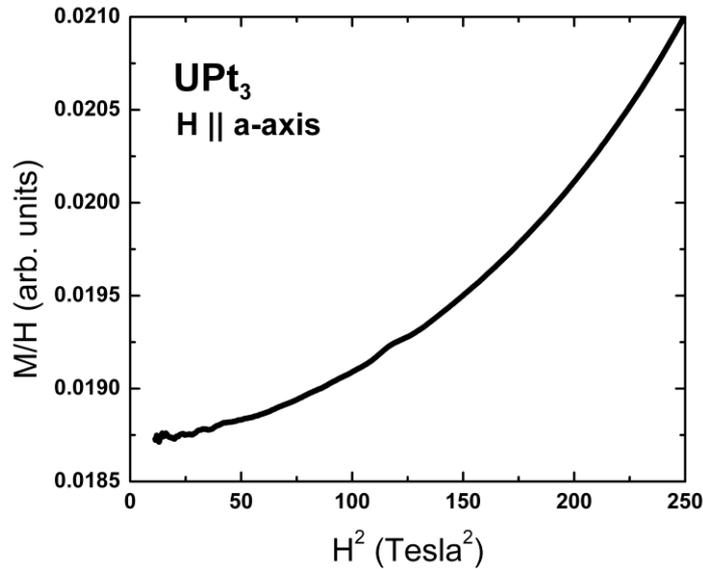

*Figure 4: Shows the result of replotting the data in fig.3 confined to within a field of 16 tesla as per eqn.(10). The deviation from a straight line is clear thus indicating a significant contribution from $\chi_5$, the fifth order susceptibility term to the magnetization.*

In conclusion, we have presented a simple scheme for extracting the higher order susceptibilities $\chi_3$ and $\chi_5$ from torque magnetometry measurements carried out for two orthogonal orientations of the field with respect to a weakly magnetic crystalline sample. Representative data on the correlated metal UPt$_3$ where $\chi_5$ is obtained for the first time through this simple scheme reveals the promise that this method holds. Complementary measurements performed at low fields on the same sample are helpful and recommended in cross-checking the data from the torque measurements.

**ACKNOWLEDGEMENTS:** We like to acknowledge the invaluable assistance of Donovan Hall and Eric Palm at the NHMFL in Florida where the measurements reported here were carried out and many useful conversations with Pradeep Kumar. We acknowledge the assistance of Brian Dorsey in acquiring the data presented in this note. The NHMFL is supported by the National Science Foundation. The work at the University of Virginia was made possible through grant NSF DMR 0073456.